\documentclass[%
 reprint,
 superscriptaddress,
 amsmath,amssymb,
 aps,
onecolumn,
pra,
 longbibliography
]{revtex4-2}
\usepackage{array}
\usepackage{lineno}
\usepackage{graphicx}
\usepackage{dcolumn}
\usepackage{bm}
\usepackage{physics}
\usepackage{booktabs}
\usepackage{textcomp}
\usepackage{bbm}
\usepackage{subfigure}
\usepackage{xcolor}

\begin{document}
\preprint{APS/123-QED}

\title{Relativistic Position Verification with Coherent States}
\author{Guan-Jie Fan-Yuan}
\thanks{These authors contributed equally}
\author{Yang-Guang Shan}
\thanks{These authors contributed equally}
\affiliation{Laboratory of Quantum Information, University of Science and Technology of China, Hefei 230026, China}
\affiliation{Anhui Province Key Laboratory of Quantum Network, University of Science and Technology of China, Hefei 230026, China}
\affiliation{CAS Center for Excellence in Quantum Information and Quantum Physics, University of Science and Technology of China, Hefei 230026, China}
\author{Cong Zhang}
\thanks{These authors contributed equally}
\affiliation{Institute of Advanced Photonics Technology, School of Information Engineering, Guangdong Provincial Key Laboratory of Information Photonics Technology, and Key Laboratory of Photonic Technology for Integrated Sensing and Communication, Ministry of Education of China, Guangdong University of Technology, Guangzhou 510006, China}
\author{Yu-Long Wang}
\author{Yu-Xuan Fan}
\author{Wei-Xin Xie}
\affiliation{Laboratory of Quantum Information, University of Science and Technology of China, Hefei 230026, China}
\affiliation{Anhui Province Key Laboratory of Quantum Network, University of Science and Technology of China, Hefei 230026, China}
\affiliation{CAS Center for Excellence in Quantum Information and Quantum Physics, University of Science and Technology of China, Hefei 230026, China}
\author{De-Yong He}
\author{Shuang Wang}
\email{wshuang@ustc.edu.cn}
\author{Zhen-Qiang Yin}
\author{Wei Chen}
\affiliation{Laboratory of Quantum Information, University of Science and Technology of China, Hefei 230026, China}
\affiliation{Anhui Province Key Laboratory of Quantum Network, University of Science and Technology of China, Hefei 230026, China}
\affiliation{CAS Center for Excellence in Quantum Information and Quantum Physics, University of Science and Technology of China, Hefei 230026, China}
\affiliation{Hefei National Laboratory, University of Science and Technology of China, Hefei 230088, China}
\author{Song-Nian Fu}
\affiliation{Institute of Advanced Photonics Technology, School of Information Engineering, Guangdong Provincial Key Laboratory of Information Photonics Technology, and Key Laboratory of Photonic Technology for Integrated Sensing and Communication, Ministry of Education of China, Guangdong University of Technology, Guangzhou 510006, China}
\author{Guang-Can Guo}
\author{Zheng-Fu Han}
\affiliation{Laboratory of Quantum Information, University of Science and Technology of China, Hefei 230026, China}
\affiliation{Anhui Province Key Laboratory of Quantum Network, University of Science and Technology of China, Hefei 230026, China}
\affiliation{CAS Center for Excellence in Quantum Information and Quantum Physics, University of Science and Technology of China, Hefei 230026, China}
\affiliation{Hefei National Laboratory, University of Science and Technology of China, Hefei 230088, China}


\begin{abstract}
Determining the position of an entity is a fundamental prerequisite for nearly all activities. Classical means, however, have been proven incapable of providing secure position verification, meaning that a prover can mislead verifiers about its actual position. In this work, we propose and experimentally realize a secure position-verification protocol that leverages quantum optics and relativity within an information-theoretic framework. Using phase-randomized weak coherent states, two verifiers separated by 2 km securely verify the prover’s position with an accuracy better than 75 meters. These results establish secure position-based authentication as a practical possibility, paving the way for applications in financial transactions, disaster response, and authenticated secure communications.

\end{abstract}

\maketitle

\section{Introduction}
Since the Internet became ubiquitous, activities such as remote transactions, online communication, and digital governance have reshaped how societies function. However, as global connectivity has accelerated, the ability to trust where information originates has not kept pace. In a world where identities, assets, and autonomous agents interact remotely, the absence of verifiable position leaves a critical gap in digital trust. Establishing whether an entity is physically present at a claimed position has therefore become a crucial security credential for identity authentication~\cite{chandran2009position,malaney2010location,kon2025quantum}.

Despite its importance, securely verifying position is fundamentally challenging. A typical position-verification protocol~\cite{bozzio2024quantum} uses multiple verifiers that send coordinated challenge messages to a prover, who must derive a credential from all received messages and return it. Because the round-trip time is constrained by the relativistic light-speed limit, the verifiers can determine the prover’s position by checking both the credential and its arrival time~\cite{brands1994distance}. However, classical position verification has been proven insecure~\cite{chandran2009position} because classical information can be copied and relayed without detection, allowing a dishonest prover to deploy collaborating adversaries that intercept and forward the messages, generate the correct credential without added delay, and thereby deceive the verifiers about its position.

Guarding against dishonest provers cannot rely solely on relativistic spacetime relations. Encoding classical information into qubits makes interception and resending detectable, forming the basis of quantum-secure communication~\cite{bennet1984quantum}. Integrating qubits into position verification therefore offers a physically grounded way to overcome the classical impossibility of secure positioning~\cite{bluhm2022a,escol2023single,allerstorfer2025making,wang2025secure,kent2011quantum,PhysRevA.84.022335,tomamichel2013monogamy}. Recent protocols~\cite{bluhm2022a,escol2023single} adopt this approach by constructing challenge messages consisting of $n$ classical bits and a single qubit, requiring adversaries to possess $O(n)$ entangled pairs to mount a successful attack. Since preparing classical bits is significantly simpler than preparing entangled states, this design satisfies the fundamental asymmetry principle of cryptography, which demands that passing the verifiers' checks be easy for an honest prover but hard for any adversary attempting impersonation.

Although conceptually appealing, qubit-based position-verification protocols are extremely challenging to realize, with no complete experimental demonstration reported to date~\cite{bozzio2024quantum,kanneworff2025experimental}. First, relativistic constraints make the verification process extraordinarily sensitive to delay. Typical millisecond-scale excess latency in modern communication translates into position errors of hundreds of kilometers at the speed of light, which eliminates practical viability of the protocol. More critically, within the information-theoretic framework, it has been proven that the prover is required to support $2^{2^{n}}$ doubly-exponential credential computations for $n>20$~\cite{escol2023single}, i.e. approximately $10^{315653}$, where larger $n$ provides stronger security. This scale imposes prohibitive computational demands and leads to substantial latency. Finally, the protocol utilizing quantum optics requires single-photon sources, high repetition rates, and a total loss not exceeding 3\,dB including modulation, transmission, and detection. Each one poses substantial practical difficulty, and satisfying them simultaneously is even more demanding.

In this paper, We realize a complete quantum position verification (QPV) by experimentally combining quantum optics with relativity within an information-theoretic framework, achieving a building-scale verification precision better than $75\,\mathrm{m}$, comparable to typical secure-zone dimensions and indicating practical relevance. In Section~\ref{sec:protocol}, we establish a security framework based on phase-randomized weak coherent states (PR-WCSs), which resolves the multiphoton security issue and eliminates the need for single-photon sources. Immediate advantages are that coherent-state sources are readily available, insensitive to modulation loss, and naturally compatible with high repetition rates. This enables a high-speed, low-loss polarization-encoding scheme, implemented using a Sagnac architecture constructed from a micro-assembled rotated circulating splitter (RCS) to achieve high-fidelity polarization-state preparation. On the detection side, we develop a high-frequency, high-voltage–driven optical switch combined with superconducting detectors to realize low-loss polarization analysis. The overall quantum-optical efficiency reaches 70\%, while maintaining an error rate of 0.27\%.

To address the latency associated with $n$-bit classical messages, we further develop dedicated classical links based on dense-wavelength-division-multiplexed on–off keying (DWDM–OOK) signal generation, anti-resonant hollow-core fiber (AR-HCF) transmission, and high-speed PIN photodiode detection. This minimal parallel transmission–detection design, together with the high-bandwidth, near-light-speed channel, enables scalable $n$ with negligible excess latency. Finally, we implement large-$n$ high-speed credential computation using a hardware lookup table built on a field-programmable gate array (FPGA) and double data rate (DDR) memory array, enabling fast mapping with a computation latency below $118\,\mathrm{ns}$ over a computation space exceeding $10^{330985980541}$, corresponding to $n=40$. Implementation details are described in Section~\ref{sec:implementation}, with further discussion presented in Section~\ref{sec:discussion}.

\section{Protocol}
\label{sec:protocol}
The mechanism of quantum position verification is that the prover performs the required operations based on the information provided by the verifiers and returns the outcome for verification. When the prover behaves honestly, the round-trip time of the information exchange can be used to bound the prover’s position. Accordingly, a quantum position-verification protocol can be organized into three components: (i) message preparation, (ii) credential generation, and (iii) position inference. In the message-preparation component, the verifiers create and send challenge messages composed of classical bits and a qubit to the prover. Credential generation then takes place at the prover, who applies the agreed rule to produce the credential and returns it immediately. Finally, after repeating these steps for $N$ rounds, the verifiers perform position inference based on the correctness of the $N$ credentials, while the largest excess latency relative to the light-speed limit across the rounds determines the uncertainty of the inferred position.

\begin{figure*}[htbp]
    \centering
    \includegraphics[width=0.98\linewidth]{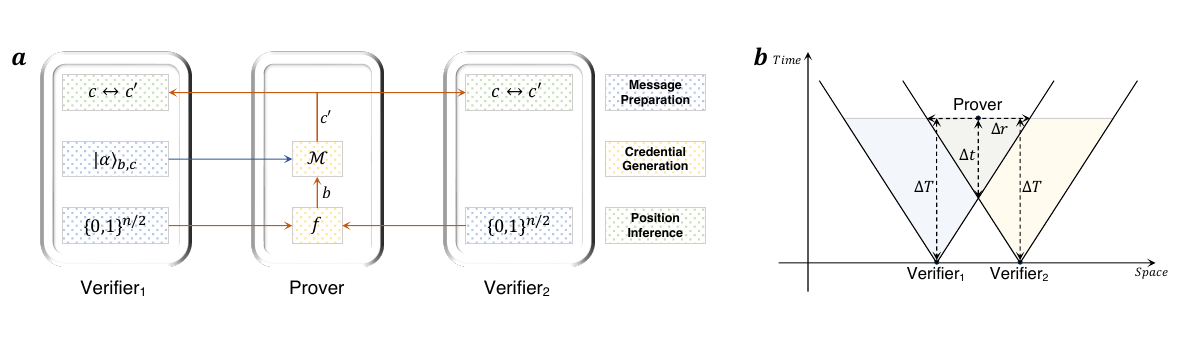}
    \caption{Quantum position verification protocol using coherent states. The position of the prover, $P$, is verified by two spatially separated verifiers, $V_1$ and $V_2$. \textbf{a.} Information flow in a single round of position verification. $V_1$ and $V_2$ each send $n/2$ classical bits to $P$. Simultaneously, $V_1$ sends a coherent state $\ket{\alpha_{b,c}}$ with mean photon number $\abs{\alpha}^2$, whose polarization is determined by eigenbasis $b$ and eigenvalue $c$. $P$ evaluates the Boolean function $f:\{0,1\}^n\to b$ to obtain $b$, measures ($\mathcal{M}$) the polarization of the quantum state accordingly, and returns the result $c'$ as a credential to the verifiers for verification. \textbf{b.} Light-cone interpretation of position precision. Without any delay, $P$ is precisely located at the intersection of the light cones from $V_1$ and $V_2$. An excess delay $\Delta t$ allows all events within a range $\Delta r$ to satisfy the verification condition, so $P$'s position is constrained within $\Delta r$.
}
    \label{fig:protocol}
\end{figure*}

Fig.~\ref{fig:protocol}a illustrates the structure of the protocol, consisting of two verifiers, $V_1$ and $V_2$, and a prover $P$, together with the flows of classical and quantum information, shown in orange and blue arrows, respectively. The verifiers jointly send $n$ classical bits to the prover, each contributing $n/2$ bits, denoted as $\{0,1\}^{n/2}$. In addition, $V_1$ sends a coherent state $\ket{\alpha_{b,c}}$ to the prover, and $b,c\in\{0,1\}$ jointly determine the polarization of the state. The bit $b$ specifies one of the two eigenbases, and $c$ specifies one of the two eigenvalues. The value of $b$ is determined through a Boolean function $f:\{0,1\}^n\to b$, which is also the operation carried out by the prover, as indicated by $f$ in the figure. The operation $\mathcal{M}$ at the prover represents a projective measurement on the received quantum state in the inferred eigenbasis, yielding the outcome $c$. This value $c$ serves as the credential returned to the verifiers for verification.

Before the protocol begins, the verifiers agree on the total number of rounds $N$, the classical string $\{0,1\}^{n}$ and the bits $b$ and $c$ for each round, and they also agree with the prover on a Boolean function. This Boolean function is selected uniformly at random from the set of all possible mappings, of which there are $2^{2^{n}}$ for input length $n$.

\textbf{Message Preparation.} For the classical part of the message, the verifiers each send the agreed-upon $\{0,1\}^{n/2}$ string for the current round and record the time at which their respective classical message enters the channel, denoted as $t_1$ and $t_2$. For the quantum part, $V_1$ prepares a weak coherent state $\ket{\alpha}$ and modulates its polarization according to $b$ and $c$, after which the state is sent to the prover.

\textbf{Credential Generation.} Upon receiving the two $\{0,1\}^{n/2}$ strings from the verifiers, the prover combines them into an $n$-bit input and applies the Boolean function to obtain $b$. The prover then measures the quantum state sent by $V_1$ in the basis specified by $b$, yielding an outcome $c'$. This value $c'$ is subsequently returned to the verifiers. Note that $V_1$ can adjust the transmission time of the quantum state so that it arrives exactly when the prover performs the measurement, thereby avoiding any need for quantum memory. Due to imperfections in state preparation and measurement, $c'$ may differ from $c$. Moreover, because of optical loss and finite detection efficiency, the prover may occasionally obtain no measurement outcome, denoted as $\perp$, so that $c' \in \{0,1,\perp\}$.

\textbf{Position Inference.} The verifiers receive the value $c'$ and record its arrival times as $t_1'$ and $t_2'$, respectively. If $c' = c$, the event is classified as a correct event. If $c' \neq c$, the event is counted as an incorrect and discard event. If the prover obtains no measurement outcome, the event is recorded as a no-response event. After $N$ rounds, the number of correct events is denoted as $n_c$, the number of incorrect and discard events as $n_I$, and the number of no-response events as $n_\perp$. The verification decision is made by setting a scoring scheme and threshold such that an honest $P$ can exceed the threshold, while a dishonest $P$ cannot. The score is calculated as $\Gamma=\gamma_Cn_c-\gamma_\perp n_\perp-\gamma_In_I$, where the coefficients for each term are obtained by semidefinite programming (SDP)\cite{escol2023single,Vandenberghe1996Semidefinite}. Let $\Gamma_0$ be the threshold. With an appropriate choice of $\Gamma_0$, the probability that a dishonest prover's score exceeds this threshold can be exponentially small, while an honest prover's score can almost certainly surpass it. Therefore, if $\Gamma \ge \Gamma_0$, $P$ is considered to have passed the verification, and its possible position region is given by the intersection of the two areas centered at the two verifiers, with radii $(t_1' - t_1)c_0/2$ and $(t_2' - t_2)c_0/2$, respectively, where $c_0$ denotes the speed of light in vacuum.

Fig.~\ref{fig:protocol}b illustrates the relationship between latency and the admissible position range using a light-cone representation. Suppose the verifiers observe a round-trip time of $2\delta T$ for receiving the credential, i.e., a one-way time of $\delta T = (t_1' - t_1)/2 = (t_2' - t_2)/2$. Under the light-speed limit, each verifier’s admissible region corresponds to the blue and orange areas within its respective light cone, and the intersection of these regions specifies where the prover could be located. If no additional latency $\delta t$ is present, the two regions become tangent, allowing the prover to be localized to a single point. Once extra latency arises, the prover’s position is instead constrained within a range of size $\delta r = c_0\delta t$.

Compared with previous protocols~\cite{bluhm2022a,escol2023single}, the key distinction of our protocol is the use of coherent states instead of single photons. Coherent states can be readily generated by laser sources and do not degrade under attenuation, making them particularly suitable for the realization of quantum position verification. However, their multiphoton components allow undetectable polarization-state cloning via beamsplitting attacks and thus introduce security concerns, while the vacuum component carries no information and reduces the verification efficiency.

To address these issues, we construct a coherent-state version of $\Gamma_0$ based on analysing different photon-number components separately and determining an upper bound on $\Gamma_0$ under adversarially optimal conditions. We employ phase-randomized coherent states, which are mixtures of Fock states with photon numbers following a Poisson distribution $\pi(\abs{\alpha}^2)$. Under this representation, each protocol round is equivalent to $V_1$ sending a polarization-encoded Fock state.

For single-photon states, we adopt the approach employed in previous protocols. For vacuum states, an adversary may declare a no-response event or may produce a correct event with probability one half. By enumerating all possibilities, we derive the vacuum-state contribution to the upper bound of $\Gamma_0$. For multiphoton states, we assume the worst case in which the adversary can always mount a perfect attack, yielding only correct events. Combining all photon-number contributions yields the desired upper bound on $\Gamma_0$.

Moreover, although the probabilities of the three photon-number cases are known, finite rounds introduce statistical fluctuations. Using the Chernoff bound, we obtain a rigorous finite-size upper bound on $\Gamma_0$. By inversely optimizing the mean photon number that maximizes the honest prover’s score, we further enhance the robustness of the verification. The analysis of the $\Gamma_0$ upper bound, together with the treatment of statistical fluctuations and coherent-state parameter optimization, is presented in detail in the Secure Score with Coherent States section of the Methods.

The protocol not only resolves the key practical bottlenecks in availability, operability, and loss tolerance through the use of coherent states, but also establishes finite-size secure bounds and introduces a parameter-optimization method that enhance the robustness. These contributions elevate quantum position verification from a theoretically defined concept to a practically realizable level and lay the foundation for subsequent experimental implementations.

\section{Implementation}
\label{sec:implementation}
\begin{figure*}[!htbp]
    \centering
    \includegraphics[width=0.98\linewidth]{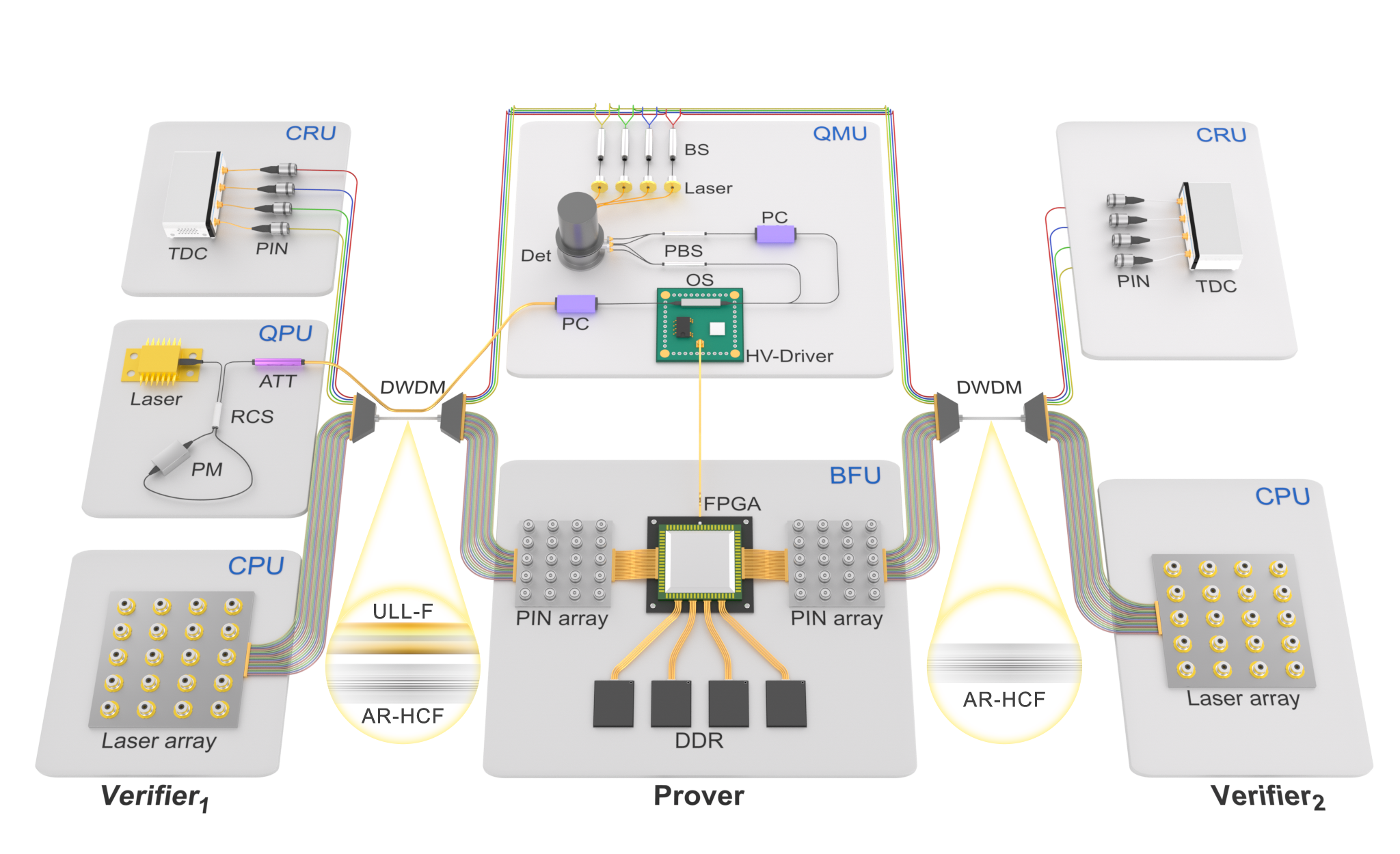}
    \caption{Experimental setup for quantum position verification. From left to right are $V_1$, $P$, and $V_2$, spaced approximately 0.98 km apart in sequence. The classical bits preparation unit (CPU) and quantum state preparation unit (QPU) of the verifiers implement the Message Preparation step of the protocol, while the credential receive unit (CRU) implements the Position Inference step. The prover's Boolean function unit (BFU) and quantum state measurement unit (QMU) implement the Credential Generation step. The quantum states are prepared in the $\ket{H}+e^{i\phi}\ket{V}$ polarizations using a Sagnac interferometer including a phase modulator (PM) and a micro-assembled rotated circulating splitter (RCS), and transmitted to $P$ through ultra-low-loss fiber (ULL-F). All classical signals are sent using a laser array in dense-wavelength-division-multiplexed on–off keying (DWDM-OOK) and received by high-speed PIN detectors, transmitted near the speed of light through anti-resonant hollow-core fiber (AR-HCF). The BFU is implemented using table lookups and logic computations on an FPGA with 1-Tb DDR memory, where the PIN array signals serve as addresses and the stored data as the output measurement basis. The QMU uses high-voltage driven a low-loss optical switch (OS) to select a measurement basis and superconducting detectors (DET) for detection. $V_1$ and $V_2$ employ time-to-digital converters (TDCs) to capture the returned measurement signals and record the latency. ATT, attenuator; PC, polorization controller; BS, beam splitter; PBS, polarizing beam splitter.
}
    \label{fig:system}
\end{figure*}

The experimental system follows the protocol architecture shown in Fig.~\ref{fig:system}. $V_1$ consists of three parts, a classical bits preparation unit (CPU) and a quantum state preparation unit (QPU) for Message Preparation, and a credential receive unit (CRU) for Position Inference. $V_2$ has the same structure as $V_1$ except it does not include a QPU. $P$ contains a Boolean function unit (BFU) and a quantum state measurement unit (QMU) for Credential Generation. In the implementation, the loss and error rate of quantum state transmission and measurement determine whether the protocol can be successfully executed, and the excess latency from classical bits transmission and prover operations affects the precision of position verification. The use of coherent states eliminates the impact of quantum state preparation loss. 

The quantum part including QPU and QMU needs to be efficient enough to meet the given security threshold. On the one hand, it is necessary to control losses and error rates to enable the honest prover to achieve a higher score. On the other hand, the bit rate needs to be increased to raise the threshold of quantum resources required for attacks. We adopt the lowest-loss solutions for each loss-sensitive component. The channel employs the ultra-low-loss fiber (ULL-F) with attenuation of 0.142 dB/km, basis selection is implemented using optical switches with 0.6 dB insertion loss, and detection is performed using superconducting single-photon detectors with 90\% efficiency. As a result, the overall system transmission efficiency reaches 70\%.

However, low-loss optical switches typically operate at only kHz frequency, which limits the bit rate. To overcome this, we developed a fast-response high-voltage driver that enables the switch to operate at 2 MHz, thereby raising the security resource threshold to $2(n/4-5)$ Mbps. Finally, to prepare quantum states with low error, we use a Sagnac-based setup to generate four polarization states of $\ket{H}+e^{i\phi}\ket{V}$, $\phi\in\{0,\pi,\pi/2,3\pi/2\}$. The beam splitter in the Sagnac is a micro-assembled rotator, circulator, and PBS, named rotated circulating splitter. This enables simple and stable state encoding with a quantum bit error rate lower than 0.27\%. The specific details are provided in the High-fidelity quantum state preparation section of the Methods.

The latency is measured from the moment a verifier sends the basis information to the moment it receives the measurement result. Ideally, the total delay should closely match the round-trip flight time at the speed of light in vacuum. Unfortunately, the transmission of classical bits and credential, the execution of the Boolean function, and the basis selection and detection of the quantum state all introduce additional delay. A more challenging aspect is that the Boolean function requires a large $n$ and the quantum measurement demands high efficiency, which often conflicts with the goal of low latency.

We primarily focus on the implementation of the Boolean function, as it directly impacts both the delay and $n$. We design a circuit based on FPGA and DDR memory to implement Boolean functions, where the basis information is used as the data address input, and the corresponding stored bit serves as the output. Different random bit sequences filling the DDR correspond to different Boolean function mappings. The total DDR capacity is $1\ \text{Tb} = 2^{40}\ \text{bits}$, which corresponds to $n = 40$. By leveraging parallelism and combinational logic, the Boolean function operates with a delay of 117.3 ns, primarily due to DDR access latency. The specific details are provided in the Large-scale and rapid Boolean function operation section of the Methods.

Next, we address the transmission delay, especially for the 40-bit basis information. By employing 978.4 m and 981.2 m AR-HCF links between the verifiers and the prover, along with DWDM-OOK encoding for simultaneous bit transmission, the excess delays from $V_1$ and $V_2$ to $P$ are significantly reduced to 22.05 ns and 22.39 ns, respectively. Then, the high-voltage driver for the low-loss optical switch is based on GaN, enabling 400 V peak-to-peak voltage switching within 50 ns. A delay of approximately 17.7 ns is introduced by the single-photon detector due to its internal wiring. The remaining delay of about 20 ns is caused by the unavoidable fiber and cable connections between components. Detailed explanations of these two parts are provided in the Near-light-speed channel and Low-delay and low-loss quantum measurement sections of the Methods.

In the experiment, a random bit sequence is loaded into the memory of the Boolean function circuit, representing the selected function. The two verifiers send verification streams to the prover at a frequency of 2 MHz, while $V_1$ simultaneously sends qubits to the prover at the same rate. The prover decodes the basis information and measures the qubit accordingly, then immediately sends the single-photon detector results back to each verifier. The number of rounds required for successful verification and the intensity of the transmitted weak coherent states are $10^7$ and 0.52, respectively, in our experiment. These parameters are optimized based on the error rate and transmittance of the quantum subsystem.

\begin{table}[!htbp]
\caption{\label{tab:result}Experimental results%
}
\begin{ruledtabular}
\begin{tabular}{cccccc}
 & Total Count & Correct Count & Error Count & No-Response Event & Score/Threshold\\
\colrule
 Theory  & 3029620 & 3020530 & 9090 & 6970380 & -242972 \\
 Trial 1 & 2985866 & 2977742 & 8124 & 7014134 & -232811.47 \\
 Trial 2 & 2984683 & 2976707 & 7976 & 7015317 & -232767.09 \\
 Trial 3 & 2987501 & 2979473 & 8028 & 7012499 & -232559.41 \\
 Trial 4 & 2980645 & 2972696 & 7949 & 7019355 & -233114.21 \\
 Trial 5 & 2982920 & 2975003 & 7917 & 7017080 & -232869.41 \\
\end{tabular}
\end{ruledtabular}
\end{table}

Table~\ref{tab:result} presents the theoretical and experimental values for each type of event, where the theoretical score corresponds to the threshold $\Gamma_0$. The results show that the score obtained in all five experimental trials exceeded $\Gamma_0$, thereby achieving successful verification. The average score of the verification system is $-232824.32\pm199.80$, significantly above $\Gamma_0$, demonstrating the robustness of the system enabled by the optimal selection of coherent states. The prover’s maximum response delay is 247.8~ns, corresponding to a verification range of 74.3~meters, achieving a level of practical significance.

\section{Conclusion}
\label{sec:discussion}
We demonstrate secure position verification grounded in both relativistic and quantum principles. Given two verifiers with known coordinates, the claimed position of a remote prover can be authenticated within 75~m of accuracy in only a few seconds, enabled by an overall system latency in the nanosecond regime. The security of the scheme relies on the light-speed limit and the non-reciprocity between classical and quantum resources. With a Boolean size of $n=40$ and a 2~MHz repetition rate, adversaries would be required to distribute no fewer than 10~M entangled qubit pairs per second with perfect fidelity, exceeding current capabilities by more than two orders of magnitude~\cite{Wengerowsky2018An,liu2019energy,Neumann2022Continuous,zhuang2025ultrabright}. This establishes a milestone in elevating quantum position verification from a theoretical concept to a practically deployable technology.

In real life, position is a fundamental element of human activity. Our method provides a solution for position-based authentication scenarios, where certain actions are permitted only when a user is physically present at a specific position. For example, allowing transactions only near an ATM, granting database access only inside an office, unlocking tickets only at a concert venue, or enabling vehicle access only beside a rental car. In addition, this method can be applied to the position tracking of critical targets, such as individuals in disaster relief, high-value goods, and military assets like weapons and ammunition.

\section*{Acknowledgments}
We cordially thank Florian Speelman, Andreas Bluhm and Llorenç Escolà-Farràs for many helpful discussions. This work was supported by the National Natural Science Foundation of China (Grant No. 62571508, 62425507, 62531023, 62271463, 62301524, 62371437), Natural Science Foundation of Anhui (No. 2308085QF216), and Quantum Science and Technology-National Science and Technology Major Project 2021ZD0300701.

\section*{Author contributions}
G.-J.F.-Y., Y.-G.S., C.Z. and S.W. conceived the idea and designed the study. Y.-G.S., Z.-Q.Y.. and W.C. developed the theoretical framework. G.-J.F.-Y., C.Z., D.-Y.H., S.W. and S.-N.F. developed the experimental methodology. G.-J.F.-Y., Y.-L.W., Y.-X.F. and W.-X.X. designed and implemented the experimental setup. G.-J.F.-Y., Y.-L.W., performed the experiments and collected the data. G.-J.F.-Y., Y.-G.S. and S.W. wrote the manuscript with input from all authors. S.W., S.-N.F., G.-C.G. and Z.-F.H. supervised the project. All authors reviewed and approved the final version of the manuscript.

\section*{Competing interests}
The authors declare no competing interests.

\section*{Data availability}
All data supporting the findings of this article are available from the corresponding authors upon request.

\appendix
\renewcommand{\thefigure}{A.\arabic{figure}} 
\setcounter{figure}{0} 
\renewcommand{\thetable}{A.\arabic{table}} 
\setcounter{table}{0} 
\renewcommand{\theequation}{A.\arabic{equation}} 
\setcounter{equation}{0} 

\section{Methods}
\subsection{Secure Score with Coherent States}
In quantum communication practice, using weak coherent states is much more convenient than employing single-photon sources, as weak coherent states can be easily prepared by simply attenuating laser pulses. This approach offers two major advantages, insensitivity to losses in the sender’s devices and the ability to achieve high repetition rates. It is crucial for quantum position verification, as the protocol is loss-sensitive and requires more than $10^6$ rounds of repetition.

To facilitate security analysis, we employ phase-randomized weak coherent states, which are mixed states of Fock states. Such sources are readily available, with phase randomization naturally achieved using gain-switched lasers. We proceed to derive $\Gamma_0$ for the case of weak coherent states, taking into account statistical fluctuations due to the finite number of rounds, thereby establishing a rigorous security bound.

PR-WCSs are mixed states of Fock states, i.e. $\int_{0}^{2\pi}\ket{\alpha e^{\text{i}\theta}}\bra{\alpha e^{\text{i}\theta}}=\sum_{i=1}^\infty e^{-\abs{\alpha}^2}\frac{\abs{\alpha}^i}{i!}\ket{i}\bra{i}$, where $\ket{\alpha e^{\text{i}\theta}}$ is the coherent state with a complex amplitude $\alpha e^{\text{i}\theta}$, and $\ket{i}$ is the $i$-photon Fock state. Therefore, each round of quantum state preparation can be categorized into three types, vacuum states, single-photon states, and multi-photon states. For single-photon states, the same approach as in Ref.\cite{escol2023single} can be applied. Vacuum states carry no information, so an adversary cannot attack them. For multi-photon states, we adopt the most pessimistic scenario, assuming that the adversary can perfectly attack them. Then, $\Gamma_0$ can be written as
\begin{equation}
    \Gamma_0=S_0^u+S_1^u+S_{2+}^u,
\end{equation}
where $S_0$, $S_1$, $S_{2+}$ represent the scores corresponding to vacuum, single-photon, and multi-photon states, respectively. The superscript $u$ denotes the upper bound that a dishonest prover can achieved. We next analyze the upper bounds separately. In the following analysis, we use probabilistic inequalities with failure probability $\epsilon$ for $5$ times. So the failure probability of the protocol is $5\epsilon$. In our experiment, we set $\epsilon=10^{-10}$, so the total failure probability is $5\times10^{-10}$. This means a dishonest prover cannot surpass the threshold $\Gamma_0$ with a probability larger than $5\times 10^{-10}$.

\subsubsection{Upper bound for vacuum states}
For vacuum states, the adversary may either declare a no-response event or a response event. A no-response contributes to $n_\perp$. A response event yields a correct outcome with 50\% probability, thereby contributing to either $n_c$ or $n_I$. Let $N_0$ be the total number of vacuum-state rounds, $x$ the number of rounds where the adversary declares a response, and the remaining $N_0-x$ rounds correspond to no-response events. Let $Y_i$ denote the score obtained by the adversary in the $i$-th round with a declared response. Then, for all rounds of vacuum states, the adversary’s total score, $S_0$, is given by
\begin{equation}
    S_0=\sum_{i=1}^x{Y_i}-\left(N_0-x\right)\gamma_\perp.
\end{equation}
For any given response round, since the vacuum state does not reveal any encoded information, each round is independently and identically distributed, yielding a correct or incorrect response with equal probability of 50\%. By normalizing $Y_i$ as a Bernoulli random variable $\frac{Y_i+\gamma_I}{\gamma_C+\gamma_I}$, and applying the Chernoff bound \cite{chernoff1952measure} for independent random variables, we obtain
\begin{equation}
    \sum_{i=1}^x{Y_i}\leq\frac{\gamma_c-\gamma_I}{2}x+\frac{\gamma_c+\gamma_I}{2}\left(\ln\frac{1}{\epsilon}+\sqrt{\ln^2\frac{1}{\epsilon}+4(\ln\frac{1}{\epsilon})x}\right),
\end{equation}
where $\epsilon$ is the failure probability. So the upper bound of $S_0$ is given by
\begin{equation}
    S_0\leq\frac{\gamma_c-\gamma_I}{2}x+\frac{\gamma_c+\gamma_I}{2}\left(\ln\frac{1}{\epsilon}+\sqrt{\ln^2\frac{1}{\epsilon}+4(\ln\frac{1}{\epsilon})x}\right)-\left(N_0-x\right)\gamma_\perp.
\end{equation}
Noting that $\max_x(ax+b\sqrt{c+dx})\to x=-\frac{c}{d}+\frac{b^2d}{4a^2}$, the score $S_0$ reaches its upper bound $S_0^u$ when $x=-\frac{1}{4}\ln\frac{1}{\epsilon}+\frac{(\gamma_c+\gamma_I)^2}{(\gamma_c-\gamma_I+2\gamma_\perp)^2}\ln\frac{1}{\epsilon}$. So $S_0^u$ is given by
\begin{equation}
    S_0\leq S_0^u=\frac{(7\gamma_c^2+4\gamma_c\gamma_\perp+4\gamma_\perp^2+10\gamma_c\gamma_I+12\gamma_\perp\gamma_I-\gamma_I^2)\ln{\frac{1}{\epsilon}}}{8(\gamma_c+2\gamma_\perp-\gamma_I)}+(\gamma_c+\gamma_I)\sqrt{\frac{(\gamma_c+\gamma_I)^2\ln^2{\frac{1}{\epsilon}}}{(\gamma_c+2\gamma_\perp-\gamma_I)^2}}-\gamma_\perp N_0.
\end{equation}
In the case where $\gamma_c+2\gamma_\perp-\gamma_I<0$, it can be further simplified to 
\begin{equation}
    S_0\leq S_0^u=-\frac{(\gamma_c-2\gamma_\perp+3\gamma_I)^2\ln{\frac{1}{\epsilon}}}{8(\gamma_c+2\gamma_\perp-\gamma_I)}-\gamma_\perp N_0. \label{eq:A_s0b}
\end{equation}

\subsubsection{Upper bound for single-photon states}
For single-photon states, we follow the same analysis as in Ref. \cite{escol2023single}, which gives the following result. For a single-photon round, if we assume that the attackers output different responses to the two verifiers with a probability less than $\xi$, then SDP can be employed to obtain a set of parameters $\{\gamma_c,\gamma_\perp,\gamma_I\}$, such that the adversary's expected score under these parameters does not exceed zero. The parameters in this work are selected using this approach. To ensure a reasonable choice of $\xi$, we note that in our experiment the honest prover always reports same responses. If the attackers adopt a strategy exceeding $\xi$ in more than $N_\xi$ rounds, the probability of finding no different-response events would be less than $(1-\xi)^{N_\xi}$. Therefore, with a failure probability of $\epsilon$, we can conclude that $N_\xi\le\ln\epsilon/\ln(1-\xi)$. Therefore, let $N_1$ be the number of single-photon rounds, then there are less than $N_\xi\le\ln\epsilon/\ln(1-\xi)$ rounds that the attack is not included in the analysis of Ref. \cite{escol2023single}. And there are more than $N_1-N_\xi\le\ln\epsilon/\ln(1-\xi)$ rounds that the expected scores of attackers are less than $0$.

To give a worst-case analysis, we assume in $N_\xi\le\ln\epsilon/\ln(1-\xi)$ rounds, the attackers can perfectly attack the system, which means the attacks always give correct responses. The corresponding score upper bound is $\gamma_cN_\xi\le\ln\epsilon/\ln(1-\xi)$. For the rest $N_1-N_\xi\le\ln\epsilon/\ln(1-\xi)$ rounds, Ref. \cite{escol2023single} has proved the sequential repetition, and Azuma's inequality \cite{azuma1967weighted} can be used to bound the score upper bound. We still set the failure probability of the Azuma's inequality to be $\epsilon$, then the score upper bound is given by $\sqrt{2\ln{\frac{1}{\epsilon}}(N_1-\frac{\ln{\epsilon}}{\ln{(1-\xi)}})}$. This formula applies under the condition that $\max\{\abs{\gamma_c},\abs{\gamma_\perp},\abs{\gamma_I}\}\le1.$

Combining the above two case, the score upper bound is given by
\begin{equation}
    S_{1}^u=\gamma_c\left\lceil\frac{\ln{\epsilon}}{\ln{(1-\xi)}}\right\rceil+\sqrt{2\ln{\frac{1}{\epsilon}}(N_1-\left\lceil\frac{\ln{\epsilon}}{\ln{(1-\xi)}}\right\rceil)}, \label{eq:A_s1b}
\end{equation}
where we added the ceiling because the number of rounds should be an integer.
\subsubsection{Upper bound for multi-photon states}
For multi-photon states, the worst case is that the adversary can perform a perfect attack, always producing correct responses. Thus, the score is given by
\begin{equation}
    S_{2+}^u=N_{2+}\gamma_c, \label{eq:A_s2b}
\end{equation}
where $N_{2+}$ is the number of rounds containing two or more photons.
\subsubsection{$\Gamma_0$ under statistical fluctuations}
The photon number distribution of PR-WCSs follows a Poisson distribution with mean photon number $\mu=\abs{\alpha}^2$. However, given a total of $N$ rounds, the actual numbers of vacuum, single-photon, and multi-photon events, denoted by $N_0$, $N_1$, and $N_{2+}$, are subject to statistical fluctuations. These fluctuations should be taken into account when computing $\Gamma_0$.

According to Eqs.\ref{eq:A_s0b}–\ref{eq:A_s2b}, the score contributed by $N_0$ is negative, while those from $N_1$ and $N_{2+}$ are positive. Thus, the upper bound of $\Gamma_0$ is given by $\Gamma_0^u = \Gamma_0(N_0^l, N_1^u, N_{2+}^u)$, where the superscripts $u$ and $l$ indicate the upper and lower bounds, respectively. These bounds can be obtained using the Chernoff bound,
\begin{align}
    & N_1\leq N_1^u= Ne^{-\mu}\mu+\frac{1}{2}\left(\ln\frac{1}{\epsilon}+\sqrt{\ln^2\frac{1}{\epsilon}+8(\ln\frac{1}{\epsilon})Ne^{-\mu}\mu}\right)\\
    & N_{2+}\le N_{2+}^u= N(1-e^{-\mu}-e^{-\mu}\mu)+\frac{1}{2}\left(\ln\frac{1}{\epsilon}+\sqrt{\ln^2\frac{1}{\epsilon}+8(\ln\frac{1}{\epsilon})N(1-e^{-\mu}-e^{-\mu}\mu)}\right)\\
    & N_0\ge N_0^l=N-N_1^u-N_{2+}^u
\end{align}

\subsubsection{Experimental optimization}
In practice, it is important to ensure that an honest prover can pass the verification despite potential misalignment in its measurement system. Let $p_e$ denote the misalignment error and $\eta$ the transmittance. The problem reduces to finding an optimal mean photon number $\mu$ that maximizes $\Gamma - \Gamma_0$, where the expected score of an honest prover is given by
\begin{equation}
    \Gamma = N\left(\gamma_c(1-e^{-\eta \mu})(1-p_e)-\gamma_\perp e^{-\eta \mu}-\gamma_I(1-e^{-\eta\mu})p_e\right).
\end{equation}
Based on the normalized parameters $\gamma_c = 0.04275$, $\gamma_\perp = 0.05019$, and $\gamma_I = 1$ obtained via semidefinite programming in Ref.\cite{escol2023single} when $\xi=0.001$, along with the experimental parameters $p_e = 0.3\%$ and $\eta = 70\%$, the average photon number is optimized to $\mu = 0.52$ for $N = 10^7$. In this case, $\Gamma_0$ is calculated as $-242972$.

\subsection{High-fidelity quantum state preparation}
The principle of polarization-state preparation is to split an initial polarization state into two components, introduce a relative phase by adjusting the phase of one component, and then recombine them orthogonally to generate the state $\ket{H}+e^{i\phi}\ket{V}$. Here we adopt a Sagnac structure. After the initial polarization state is divided into two parts, they propagate clockwise and counterclockwise in the Sagnac loop. A phase modulator placed at an off-center position selectively modulates the phase of only one propagation direction, thereby introducing a relative phase difference.

The key advantage of this structure is that the clockwise and counterclockwise pulses traverse the same fiber, so their relative phase does not drift due to path-length differences. This yields high stability and is favorable for high-fidelity quantum state preparation. However, conventional Sagnac schemes based on beam splitters cannot be used for polarization-state preparation because they do not support orthogonal recombination. Using a PBS would require rotating the initial polarization by $45^\circ$ with a polarization controller, which compromises stability.

To address this limitation, we design a micro-assembled rotated circulating splitter that integrates a rotator, a circulator, and a PBS. The micro-assembly process enables precise polarization rotation and accurate $50{:}50$ splitting, while the inclusion of the circulator suppresses reflections associated with the Sagnac loop. As a result, this design achieves high-fidelity polarization-state preparation with a measured fidelity of 99.73\%. The correspondingly low error rate allows the protocol to tolerate higher losses and enables the prover to attain higher scores, which constitutes one of the key factors for the successful operation of the system.

\subsection{Large-scale and rapid Boolean function}
To achieve large-scale and rapid Boolean function computation, we employ a lookup-table approach implemented with a custom FPGA and DDR-based circuit. While fast computation typically relies on logic gates, we use a lookup-table method because logic gates are impractical in terms of both resource consumption and delay. On one hand, the state space of Boolean functions is $2^{2^n}$. Assuming each logic gate can represent $m$ states, the required number of gates is $\log_m(2^{2^n})$, which grows exponentially with $n$ and far exceeds the capacity of current FPGAs. On the other hand, logic gates inevitably introduce circuit delays. When multiplied by an exponential number of gates, the resulting delay becomes prohibitive. The lookup-table approach transforms the scalability challenge into a memory-space problem. We implement the Boolean function using DDR chips with a total capacity of $1 \text{Tb} = 2^{40} \text{bit}$, corresponding to $n = 40$. Once the FPGA receives 40 bits, it uses them as an address to access the corresponding bit stored in the DDR memory and outputs the result as the basis selection signal. By writing different random bit patterns into the DDR, each configuration corresponds to a distinct Boolean function, enabling support for the full set of $2^{2^{40}}$ possible Boolean functions. Due to the inherent delay of the FPGA and the random-access performance of the DDR memory, this part introduces a delay of approximately 117.3 ns.

\subsection{Near-light-speed channel}
Both the basis information and the measurement results are classical bits and can be transmitted using the same method. The main difference is that the basis information contains $n/2$ bits, while the measurement result is a single bit. To enable rapid transmission of classical information, we adopted a simple and efficient on-off keying scheme, where strong and weak optical intensities represent 1 and 0, respectively. A further challenge is that if each bit is transmitted with a cycle duration of $\tau$, sending $n/2$ bits will introduce a delay of $(n/2 - 1)\tau$, which is unfavorable for Boolean functions with large $n$. We use dense wavelength division multiplexing with independent lasers at different wavelengths to send specific bits simultaneously. This enables the parallel transmission of $n/2$ bits, eliminating delay dependence on $n$. The optical signals are detected using high-speed PIN photodiodes and directly fed into subsequent circuits for discrimination. Signal transmission through the channel also takes time. To minimize this delay, we use anti-resonant hollow-core fiber with an air-filled core, enabling light to propagate at nearly the speed of light in vacuum. Combined with low dispersion, the additional delay is limited to approximately 22 ns.

\subsection{Low-delay and low-loss quantum measurement}
The loss in quantum state measurement primarily originates from the insertion loss of the basis selection device and the detection efficiency of the photon detector. To mitigate this, we employ low-loss optical switches and superconducting nanowire single-photon detectors. In addition, polarization-maintaining fusion splicing is used to reduce connector loss. As a result, the overall insertion loss is approximately 0.96 dB, mainly due to the insertion loss of the optical switch (about 0.6 dB). The detection efficiency reaches 90\%, leading a transmission efficiency of 72\%.

However, this configuration introduces delay challenges. High-speed, low-loss optical switches require a driving voltage exceeding 400 V and cannot directly interface with the digital outputs of the Boolean function circuit. Additionally, the weak click signal from the detectors cannot be directly transmitted over long distances to the verifiers. Signal conversion inevitably introduces delay. To address this, we develop a GaN-based high-voltage driver that converts digital signals into high-voltage signals within 50 ns, enabling fast response of the optical switch. In parallel, we design an integrated signal processing unit for the detectors that performs amplification, discrimination, and OOK encoding to reduce the return delay of detection signals. These measures lead to a significant reduction in delay, without compromising low loss and high repetition rate. 

\bibliography{apssamp}

\end{document}